\documentclass[prc,preprint,nofootinbib,superscriptaddress,showpacs]{revtex4}
\usepackage[dvips]{graphicx}
\DeclareGraphicsExtensions{.ps,.eps,.pdf,.jpg}
\usepackage{amssymb, latexsym, amsmath, amsfonts}
\usepackage{tabularx}
\usepackage{url}
\usepackage{color}

\allowdisplaybreaks

\newcommand{\mb}{\mathbf}
\newcommand{\mc}{\mathcal}
\newcommand{\taun}{\tau_{n}}
\newcommand{\taup}{\tau_{p}}
\newcommand{\rhon}{\rho_n}
\newcommand{\rhop}{\rho_p}

\newcommand{\unifi}{u n_{i} f_{i}}
\newcommand{\su}{s(u)}
\newcommand{\sigx}{\sigma (x)}
\newcommand{\mus}{\mu_{s}}
\newcommand{\nun}{\nu_{n}}
\newcommand{\rn}{r_{N}}
\newcommand{\uni}{u n_{i}}
\newcommand{\unixi}{u n_{i} x_{i}}
\newcommand{\nii}{n_{i}}
\newcommand{\cu}{c(u)}
\newcommand{\xii}{x_{i}}
\newcommand{\la}{\lambda_{1}}
\newcommand{\lb}{\lambda_{2}}

\newcommand{\spr}{s^{\prime}}
\newcommand{\cpr}{c^{\prime}}
\newcommand{\muhi}{\hat{\mu}_{i}}
\newcommand{\muh}{\hat{\mu}}
\newcommand{\mun}{\mu_n}

\newcommand{\muni}{\mu_{ni}}
\newcommand{\muno}{\mu_{no}}

\newcommand{\pd}{\partial}
\newcommand{\fii}{f_i}
\newcommand{\cald}{\mathcal{D}}
\newcommand{\caldp}{\mathcal{D}^{\prime}}
\newcommand{\nno}{n_{no}}
\newcommand{\foo}{f_{o}}

\begin{document}

\title{Kaon Condensation in Neutron Stars
with Skyrme-Hartree-Fock Models}

\author{Yeunhwan Lim}
\email{ylim9057@ibs.re.kr}
\affiliation{Department of Physics Education,
Daegu University, Gyeongsan 712-714, Republic of Korea}
\affiliation{Rare Isotope Science Project,
Institute for Basic Science,
Daejeon 305-811, Republic of Korea}

\author{Kyujin Kwak}
\email{kkwak@unist.ac.kr }
\affiliation{School of Natural Science,
Ulsan National Institute of Science and Technology (UNIST), Ulsan 689-798, Republic of Korea}

\author{Chang Ho Hyun}
\email{hch@daegu.ac.kr}
\affiliation{Department of Physics Education,
Daegu University, Gyeongsan 712-714, Republic of Korea}

\author{Chang-Hwan Lee}
\email{clee@pusan.ac.kr}
\affiliation{Department of Physics,
Pusan National University, Busan 609-735, Republic of Korea}
\affiliation{Department of Physics and Astronomy, 
State University of New York at Stony Brook, Stony Brook, NY 11794, USA}
\date{May 13, 2014}



\begin{abstract}
%
%
We investigate nuclear matter equations of state in
neutron star with kaon condensation. It is generally known that the
existence of kaons in neutron star makes the equation of state soft 
so that the maximum mass of neutron star is not likely to be greater than 
2.0 $M_{\odot}$, the maximum mass constrained by current observations. 
With existing Skyrme force model 
parameters, we calculate nuclear equations of state and check
 the possibility of kaon condensation in 
 the core of neutron stars. 
 The results show that
even with the kaon condensation, the nuclear equation of state
satisfies both the maximum mass and the allowed ranges of mass and radius.
%
\end{abstract}

\maketitle
\newpage

\section{Introduction}

Theory of nuclear matter has been tested using the properties of 
the observed nuclei whose number amounts to approximately 3,000 to date.
The Skyrme-Hartree-Fock (SHF) models have been widely used
to describe the general properties of nuclear medium and heavy nuclei in the
non-relativistic limit.
However, depending on the selection of the data and the methods to fit the model parameters,
there are now more than 100 SHF models, and new models are still in birth
due to the continuous update of the data.
Although most of the SHF models even with different model parameters can explain the 
properties of numerous known nuclei consistently, predictions on the properties 
of the infinite nuclear matter strongly depend on the models, 
especially at high densities 
far above the nuclear 
saturation density $\rho_0$ ($\sim 0.16\, {\rm fm}^{-3}$) \cite{brown2000}.
As a result, the maximum mass of stable neutron stars calculated 
with known SHF models ranges from $1.4 M_\odot$ to $2.5 M_\odot$ 
where $M_\odot$ is the solar mass \cite{thesis2012}.

At the core of neutron stars, there can be significant contributions from the exotic
states,  such as strangeness condensates, meson condensates, strange quark matter, etc., which are quite uncertain.
The effect of strange particles, such as hyperons, on the nuclear matter equation of state (EoS) 
has  been studied within the SHF models  \cite{mornas05,guleria12}.
%
%
In these works the existence of hyperons softens the nuclear matter EoS substantially 
and as a consequence the 
maximum mass of the neutron star decreases. Similar
conclusion for the effect of hyperons has been drawn from the calculations done with 
the relativistic mean field models (for example, see \cite{ryu11}).

Recently $2 M_\odot$ neutron stars in neutron star-white dwarf binaries were observed in pulsars PSR J1614$-$2230 \cite{Demo10} and PSR J0348$+$0432 \cite{Anto13}.
This implies that any realistic EoS for the stable neutron star 
should be able to explain masses
equal to or more than these values.
With this criterion, 
SHF models predicting maximum masses to be less than $2 M_\odot$ 
can be excluded from the candidates for 
the realistic models of high-density nuclear matter.

In this work, we revisit the kaon condensation and investigate
its effect on the EoS of neutron star matter.
In general, the EoS is very sensitive to the interactions of 
kaons in nuclear medium \cite{ryu07}.
Since general SHF models do not include inherent kaon interactions in them, 
we need to import kaon interactions from other theories or models.
In this work, we consider the SU(3) non-linear chiral effective model with kaons.
We 
investigate how the parameters in SHF and kaon interaction model affect 
the mass and radius of neutron stars, 
and constrain the parameter space by comparing
our results with observed neutron star masses, 
$(1.97\pm0.04)M_\odot$ of PSR J1614$-$2230 and $(2.01\pm 0.04)M_\odot$ of PSR J0348$+$0432. 

This paper is organized as follows.
In Sect.~\ref{sec2}, we describe the SHF models that we choose and the SU(3) non-linear 
chiral model for the interactions of kaons. In the same section, we derive 
basic equations from which the EoS within neutron stars is calculated. 
In Sect.~\ref{sec3}, we present our results on EoS, particle fractions, and mass-radius 
relations of neutron stars. Our conclusion and discussion are given in Sect.~\ref{sec4}.

\section{Models \label{sec2}}

\subsection{SHF models \label{sec2-1}}

The general Skyrme force model is used to generate energy density 
functional (EDF), in which the effective two body force between nucleons is introduced.
Because EDFs have been quite successful in explaining the properties of finite heavy nuclei, 
they have been also applied to the infinite dense system such as the interior of neutron stars. 
In the Skyrme type potential model, the effective interaction is given by
\begin{equation}
\begin{aligned}
v_{ij}  = & t_0 (1+x_0P_{\sigma})\delta(\mb{r}_{i}-\mb{r}_{j}) 
+\frac{1}{2}t_1(1+x_1 P_{\sigma})\frac{1}{\hbar^2}
\big[ \mb{p}_{ij}^2\delta(\mb{r}_{i}-\mb{r}_{j})
 + \delta(\mb{r}_{i}-\mb{r}_{j})\mb{p}_{ij}^2 \big] \\[7pt]
 & + t_2(1+x_2 P_{\sigma})\frac{1}{\hbar^2}\mb{p}_{ij}
 \cdot \delta(\mb{r}_i -\mb{r}_j) \mb{p}_{ij} 
 + \frac{1}{6}t_3 (1+x_3 P_{\sigma}) \rho^\epsilon (\mb{r}) 
 \delta(\mb{r}_{i} -\mb{r}_{j}) \\[7pt]
 & + \frac{i}{\hbar^2}W_{0}\mb{p}_{ij}\cdot\delta(\mb{r}_i -\mb{r}_j)
 (\boldsymbol{\sigma}_i + \boldsymbol{\sigma}_j)\times \mb{p}_{ij}
\,,
\end{aligned}
\end{equation}
where $\mb{r}=(\mb{r}_i + \mb{r}_j)/2$, 
$\mb{p}_{ij} =-i \hbar(\boldsymbol{\nabla}_i -\boldsymbol{\nabla}_j)/2$, 
$P_{\sigma}$ is the spin-exchange operator, and 
$\rho(\mb{r}) = \rho_n(\mb{r}) + \rho_p(\mb{r})$. 
Using the Skyrme force model,
the Hamiltonian density of nuclei can be written as \cite{sple2005}
\begin{equation}
\mathcal{H}_N = \mathcal{H}_{B} + \mathcal{H}_{g}
 + \mathcal{H}_{C} + \mathcal{H}_{J} .
\end{equation}
The bulk Hamiltonian density is given by
\begin{equation}
\begin{aligned}
\mathcal{H}_{B} & = 
\frac{\hbar^2}{2M}\tau_{n} + \frac{\hbar^2}{2M}\tau_{p}
+ \rho(\tau_n + \tau_p) 
\biggl[ \frac{t_1}{4}\Bigl(1+\frac{x_1}{2}\Bigr)
+ \frac{t_2}{4}\Bigl(1+\frac{x_2}{2}\Bigr)\biggr]\\[7pt]
& \quad + (\taun\rhon +\taup\rhop)
\biggl[ \frac{t_2}{4}\Bigl(\frac{1}{2}+ x_{2}\Bigr)
- \frac{t_1}{4}\Bigl(\frac{1}{2} + x_{1}\Bigr)\biggr]\\[7pt]
& \quad + \frac{t_0}{2}
\biggl[\Bigl(1+ \frac{x_0}{2}\Bigr)\rho^2
- \Bigl(\frac{1}{2} + x_{0}\Bigr)(\rhon^2+\rhop^2)\biggr]\\[7pt]
& \quad + \frac{t_3}{12}
\biggl[\Bigl(1+ \frac{x_3}{2}\Bigr)\rho^2
- \Bigl(\frac{1}{2} + x_{3}\Bigr)(\rhon^2+\rhop^2)\biggr]
\rho^{\epsilon}\,.
\end{aligned}
\label{eq:hamilN}
\end{equation}
The gradient Hamiltonian density takes the form of
\begin{equation}
\mc{H}_{g} = \frac{1}{2}Q_{nn}(\boldsymbol{\nabla}\rhon)^2
 + Q_{np}\boldsymbol{\nabla}\rhon \cdot \boldsymbol{\nabla}\rhop
 + \frac{1}{2}Q_{pp}(\boldsymbol{\nabla}\rhop)^2 \,,
\end{equation}
with 
\begin{equation}\label{eq:qq}
\begin{aligned}
Q_{nn} & =Q_{pp} =\frac{3}{16}\bigl[t_1(1-x_1) -t_2(1+x_2)\bigr] \,,\\
Q_{np} & = \frac{1}{8}\biggl[
3t_1\Bigl(1 +\frac{x_1}{2}\Bigr) -t_2 \Bigl(1+\frac{x_2}{2}\Bigr)
\biggr]\,.
\end{aligned}
\end{equation}
The Coulomb energy density is given by
\begin{equation}
\mc{H}_{C} = \frac{e^2}{2}\rhon(\mb{r})\int
\, d^{3}r^{\prime} \frac{\rhop(\mb{r}^{\prime})}
{|\mb{r}-\mb{r}^{\prime}|}
-\frac{3e^2}{4}\biggl(\frac{3}{\pi}\biggr)^{1/3}\rhop^{4/3}(\mb{r})\,,
\end{equation}
and $H_{J}$ comes from the spin-orbit interaction and is given by
\begin{equation}
\begin{aligned}
\mc{H}_{J} & = -\frac{W_0}{2}(\rhon \boldsymbol{\nabla}\cdot \mb{J}_{n}
+ \rhop\boldsymbol{\nabla}\cdot \mb{J}_{p} 
+ \rho\boldsymbol{\nabla}\cdot\mb{J}) \\ 
& \quad + \frac{t_1}{16}(\mb{J}_{n}^{2} + \mb{J}_{p}^2 - x_1 \mb{J}^2)
-\frac{t_2}{16}(\mb{J}_{n}^{2} + \mb{J}_{p}^{2} + x_2 \mb{J}^2)\,,
\end{aligned}
\end{equation}
where $\mb{J}_{n(p)} = \sum_{i} \psi_{i, n(p)}^{\dagger}\boldsymbol{\sigma} \times
\boldsymbol{\nabla}\psi_{i,n}$
is the neutron (proton) spin-orbit density, and $\mb{J} = \mb{J}_n +\mb{J}_p$.

In general, ten independent parameters
($x_{i=0,1,2,3}$,   $t_{i=0,1,2,3}$,  $\epsilon$, $W_0$)
in Hamiltonian density are fixed by the properties of
finite nuclei \cite{klmn2010}. 
With the ten parameters fixed, we calculate the nuclear matter properties 
of the infinite nuclear matter (such as neutron stars) which determine the 
maximum mass of neutron stars. In this work, we employ four SHF models 
all of which predict the maximum mass of neutron stars larger than $2 M_\odot$ 
while each model shows distinct characteristics for 
the symmetric nuclear matter properties and the stiffness of the EoS.

\begin{table}
\begin{center}
\begin{tabular}{lccccccc}
\hline
Model & ~~~~$\rho_0$ ~~~~ & ~~~ $B$ ~~~ & 
~~~ $S_v$ ~~~ & ~~~ $L$ ~~~ & ~~~ $K$ ~~~ & 
~$m^*_N/m_N$ ~ & $M_{\rm max}/M_\odot$ \\ \hline
SLy4 & 0.160 & 16.0 & 32.0 & 45.9 & 230 & 0.694 & 2.07 \\ 
SkI4 & 0.160 & 16.0 & 29.5 & 60.4 & 248 & 0.649 & 2.19 \\ 
SGI  & 0.155 & 15.9 & 28.3 & 63.9 & 262 & 0.608 & 2.25 \\ 
SV   & 0.155 & 16.1 & 32.8 & 96.1 & 306 & 0.383 & 2.44 \\ \hline
\end{tabular}
\end{center}
\caption{Nuclear matter properties and the maximum mass of neutron star 
calculated from four SHF models that we select. 
$\rho_0$: saturation density in unit of fm$^{-3}$, 
$B$: binding energy of the symmetric nuclear matter in unit of MeV, 
$S_v$: symmetry energy at the saturation density in unit of MeV, 
$L$: slope of symmetry energy at the saturation density in unit of MeV, 
$K$: compression modulus of the symmetric matter at the
saturation density in unit of MeV,
$m^*_N/m_N$: ratio of the effective mass of the 
nucleon at the saturation density ($m^*_N$) to the free mass of the nucleon ($m_N$), 
and $M_{\rm max}/M_\odot$: maximum mass of neutron star in unit of the solar 
mass ($M_\odot$).}
\label{tab1}
\end{table}

Table~\ref{tab1} summarizes nuclear matter properties and the maximum
mass of neutron stars obtained from the four SHF models.
For the four models that we select, the basic saturation properties $\rho_0$ and $B$ 
are almost identical or similar to each other, but the values of $S_v$, $L$ and $K$ vary 
significantly from model to model even though they are in the range of general acceptance.
Note that 
larger $K$ causes stiffer EoS. The table confirms that the maximum
mass of neutron star increases with $K$.


\subsection{Kaon interactions \label{sec2-2}}

Several models describe the interaction of kaon 
in nuclear medium. As for two examples, kaonic optical potential treats 
kaon interaction phenomenologically 
and the meson-exchange model mediates the interaction of the kaon 
with the background nuclear matter \cite{knorren95}.
In this work, we employ a SU(3) non-linear chiral model which was first proposed by
Kaplan and Nelson \cite{kaplan86}.
The effective chiral Lagrangian density is given as
\begin{eqnarray}
{\cal L} &=& \frac{f^2_\pi}{4} {\rm Tr} \partial_\mu U \partial^\mu U^\dagger
+ c {\rm Tr} [ m_q (U + U^\dagger -2 )] 
+ i {\rm Tr} \bar{B} \gamma^\mu  \partial_\mu B \nonumber \\ &&
+ i {\rm Tr} B^\dagger [V_0,\, B] 
- D {\rm Tr} B^\dagger \boldsymbol{\sigma} \cdot \{ {\mb A},\, B\}
-F {\rm Tr} B^\dagger \boldsymbol{\sigma} \cdot [\mb{A},\, B]  \nonumber \\  && 
+ a_1 {\rm Tr} B^\dagger (\xi m_q \xi + {\rm h.c.})B
+ a_2 {\rm Tr} B^\dagger B (\xi m_q \xi + {\rm h.c.})
+ a_3 {\rm Tr} B^\dagger B {\rm Tr}(m_q U + {\rm h.c.}),
\label{eq:knlag}
\end{eqnarray}
where $f_\pi$ is the pion decay constant ($\simeq 93$ MeV). 
Chiral fields $U$ and $\xi$ are defined by
\begin{equation}
U = \xi^2 = \exp( \sqrt{2} i M/f_\pi),
\label{eq:chiralfield}
\end{equation}
and the mesonic vector and axial currents read
\begin{equation}
V_\mu = \frac{1}{2}\left( \xi^\dagger \partial_\mu \xi
+ \xi \partial_\mu \xi^\dagger\right),\,\,
A_\mu = \frac{i}{2}\left( \xi^\dagger \partial_\mu \xi
- \xi \partial_\mu \xi^\dagger\right).
\end{equation}
The meson and baryon octet fields $M$ and $B$ are defined as
\begin{eqnarray}
M &=& \left(
\begin{array}{ccc}
\frac{1}{\sqrt{2}} \pi^0 + \frac{1}{\sqrt{6}}\eta & \pi^+ & K^+ \\
\pi^- & -\frac{1}{\sqrt{2}} \pi^0 + \frac{1}{\sqrt{6}}\eta & K^0 \\
K^- & \bar{K}^0 & -\sqrt{\frac{2}{3}} \eta
\end{array}
\right), \\
B &=& \left(
\begin{array}{ccc}
\frac{1}{\sqrt{2}} \Sigma^0 + \frac{1}{\sqrt{6}} \Lambda & \Sigma^+ & p \\
\Sigma^- & -\frac{1}{\sqrt{2}} \Sigma^0 + \frac{1}{\sqrt{6}} \Lambda & n \\
\Xi^- & \Xi^0 & -\sqrt{\frac{2}{3}} \Lambda
\end{array}
\right),
\end{eqnarray}
and $m_q$ is the quark mass matrix
\begin{equation}
m_q = \left(
\begin{array}{ccc}
0 & 0 & 0 \\
0 & 0 & 0 \\
0 & 0 & m_s
\end{array}
\right),
\end{equation}
where we assume massless up and down quarks ($m_u = m_d = 0$) and $m_s$ is the finite current mass of strange quark.
By expanding $U$ in terms of meson fields, we can obtain the kinetic energy and mass of meson fields 
which correspond to the first and second term, respectively in the first line of Eq.~(\ref{eq:knlag}). 
Note that there are other interaction terms with higher order in the meson fields due to the SU(3) symmetry.
The constant $c$ in the mass term can be determined from the relation $m^2_K = 2 c\; m_s /f^2_\pi$.
The third term in the first line of Eq.~(\ref{eq:knlag}) represents the kinetic energy of octet baryons.
The second and third lines in Eq.~(\ref{eq:knlag}) represent the interactions among mesons and baryons. 
We use $F=0.44$ and $D=0.81$ which are fixed by weak nucleon and semileptonic hyperon decays.
For $a_1 m_s$ and $a_2 m_s$, we quote the values given 
in Ref.~\cite{thorsson94}, where 
\begin{eqnarray}
a_1 m_s &=& -67\, {\rm MeV}, \\
a_2 m_s &=& 134\, {\rm MeV}.
\end{eqnarray}
In principle, the value of $a_3$ can be fixed by using the strangeness content of the proton $\langle\bar{s}s\rangle_p$
or the kaon-nucleon sigma term $\Sigma^{KN}$;
\begin{eqnarray}
m_s \langle \bar{s} s \rangle_p&=&- 2 (a_2+a_3)m_s, \\
\Sigma^{KN} &=& -\frac{1}{2} (a_1 + 2 a_2  + 4 a_3) m_s.
\end{eqnarray}
However, due to the uncertainties in these quantities, we choose four different values of 
$a_3 m_s$,  $-134$, $-178$, $-222$ and $-310$ MeV,
which correspond to the strangeness content
$\langle\bar{s}s\rangle_q = 0$, $0.05$, $0.1$ and $0.2$, respectively.

The amount of kaon condensation can be determined from local flavor changing
$\beta$-equilibrium, e.g., $n \leftrightarrow p + K^-$. This chemical
equilibrium implies $\mu_n = \mu_p + \mu_{K^-}$, where $\mu_i$ denotes
the chemical potential of particle `$i$'. For the simplicity, we use $\mu_K
= \mu_{K^-}$ afterward. Other equilibrium conditions will be discussed in the next subsection.
With the s-wave interactions only, the kaon condensate can be characterized
by the expectation value \cite{baym73}
\begin{equation}
\langle K^- \rangle = v_K e^{- i \mu_{K} t},
\end{equation}
where the amplitude $v_K$ determines the magnitude of the condensate.
Since the kaon field appears non-linear in Eq.~(\ref{eq:chiralfield}),
it is convenient
 to introduce a new parameter $\theta$ by
\begin{equation}
\theta \equiv \sqrt{2}\, \frac{v_K}{f_\pi}.
\end{equation}
Expanding Eq.~(\ref{eq:knlag}) in terms of meson and baryon fields 
and retaining the terms relevant to kaons and nucleons, we obtain
the Lagrangian for the kaon and kaon-nucleon interactions as
\begin{eqnarray}
{\cal L}_K &=& f^2_\pi \frac{\mu_K^2}{2}\sin^2\theta 
- 2m^2_K f^2_\pi\sin^2\frac{\theta}{2} \nonumber \\ & &
- n^\dagger n\left[-\mu_K + (2a_2 + 4 a_3)m_s\right]\sin^2\frac{\theta}{2} \label{eq:kaon} \\ & &
- p^\dagger p\left[-2\mu_K + (2a_1+2a_2+4a_3)m_s\right]\sin^2\frac{\theta}{2}. \nonumber
\end{eqnarray}
Taking $\mu_K$ as a Lagrangian multiplier to account for the
charge neutrality condition of the neutron star matter,
Hamiltonian density is derived from the kaon Lagrangian which is given by
Eq.~(\ref{eq:kaon}). 
We obtain the Hamiltonian density for the kaon as
\begin{eqnarray}
{\cal H}_K &=&-f^2_\pi\frac{\mu^2_K}{2}\sin^2\theta
+ 2m^2_K f^2_\pi \sin^2\frac{\theta}{2} + \mu_K \rho_p \nonumber \\ &&
-\mu_K (\rho+\rho_p) \sin^2\frac{\theta}{2} 
+ a_{K1}\rho_p\sin^2\frac{\theta}{2}
+ a_{K2}\rho \sin^2\frac{\theta}{2},
\label{eq:hamilK}
\end{eqnarray}
where 
$a_{K1} = 2 a_1 m_s$ and  $a_{K2} = (2 a_2+4a_3)m_s$.

In addition to the hadron parts discussed so far, leptonic terms
should be added for the complete description of EoS.
We consider both electrons and muons in this work, and their
Hamiltonian densities are given as
\begin{eqnarray}
\tilde{{\cal H}}_e &=& \frac{\mu^4_e}{4\pi^2} - \mu_e \rho_e, \\
\tilde{{\cal H}}_\mu &=& H(|\mu_\mu|-m_\mu) \left\{ \frac{m^4_\mu}{8 \pi^2} \left[
(2 x_\mu^2+1)x_\mu\sqrt{x^2_\mu+1} - \ln\left(x_\mu+\sqrt{x^2_\mu+1}\right)
\right] - \mu_\mu \rho_\mu \right\},
\label{eq:hamilL}
\end{eqnarray}
where $\rho_e=\mu_e^3/3\pi^2$, $H$ is the Heaviside step function, $\rho_{\mu}=k_{\mu}^3/3\pi^2$, $\mu_{\mu} = \sqrt{k^2_{\mu} + m^2_{\mu}}$, and $x_\mu = k_\mu/m_\mu$ with the Fermi momentum of muon $k_{\mu}$. 
Note that we allow negative values of $\mu_{e,\mu}$ in order to take into account the contributions of $e^+$ and $\mu^+$.

\subsection{Equilibrium conditions}\label{sec3:con}


As mentioned earlier, ten parameters in the nucleon Hamiltonian Eq.~(\ref{eq:hamilN}) are determined 
by fitting to the properties of finite nuclei for a given density $\rho$. After determining the 
ten parameters, we have only one independent variable $\rho_p$ with the constraint $\rho = \rho_n + \rho_p$.
In the following, we use $x\equiv \rho_p/\rho$ for the simplicity. In this work, 
we assume that only proton and neutron contribute to the baryon density even at the high-density core of neutron star.
In the Hamiltonians for kaons and leptons, Eqs.~(\ref{eq:hamilK}) to (\ref{eq:hamilL}), there are four independent variables
$\mu_K$, $v_K$ (or $\theta$),  $\mu_e$, and $\mu_\mu$.
By neglecting contributions from (anti)neutrinos which leave the system, we can set $\mu_K=\mu_\mu=\mu_e$. 
As a result, we have only three independent variables $x$, $\mu_e$, and $\theta$ which are determined by minimizing 
the total energy density given as 
%
%
%
\begin{eqnarray}
\epsilon = \epsilon_N + \epsilon_K + \epsilon_e + \epsilon_\mu . 
\label{eq:enden}
\end{eqnarray}
Three coupled equations are solved for $x$, $\mu_e$, and $\theta$,  
\begin{eqnarray}
\frac{\partial \epsilon}{\partial x} = 0 &:& \mu_e = 
(\mu_n^{\rm SHF} - \mu_p^{\rm SHF}) \sec^2\frac{\theta}{2} - a_{K1} \tan^2\frac{\theta}{2}, \label{eq:beta}\\
\frac{\partial \epsilon}{\partial \mu_e} = 0 &:& 
f^2_\pi \mu^2_e \sin^2\theta + (\rho+\rho_p)\sin^2\frac{\theta}{2} = 
\rho_p-\rho_e - \rho_\mu, \label{eq:cn}\\
\frac{\partial \epsilon}{\partial \theta} = 0 &:&
-\mu^2_e\cos\theta + m^2_K -\frac{1}{2 f^2_\pi} \mu_e(\rho+\rho_p)
+\frac{a_{K1}}{2f^2_\pi}\rho_p+\frac{a_{K2}}{2f^2_\pi}\rho=0, \label{eq:vk}
\end{eqnarray}
where $\mu^{\rm SHF}_{n,p}$ are the chemical potential of nucleon in the general SHF model only.
Eqs.~(\ref{eq:beta}) and (\ref{eq:cn}) are the beta equilibrium and local charge neutrality condition, respectively with kaons included.
Note that Eq.~(\ref{eq:vk}) is valid only for the densities beyond the critical density where $\theta\neq 0$.
Once these equations are solved, we can calculate the pressure
from the thermodynamic relation
\begin{equation}
p = \rho^2 \frac{\partial}{\partial \rho}\left(\frac{\epsilon}{\rho}\right),
\label{eq:pressure}
\end{equation}
and determine the EoS as a function of density $\rho$.

\begin{figure}
\begin{center}
\includegraphics[scale=0.85]{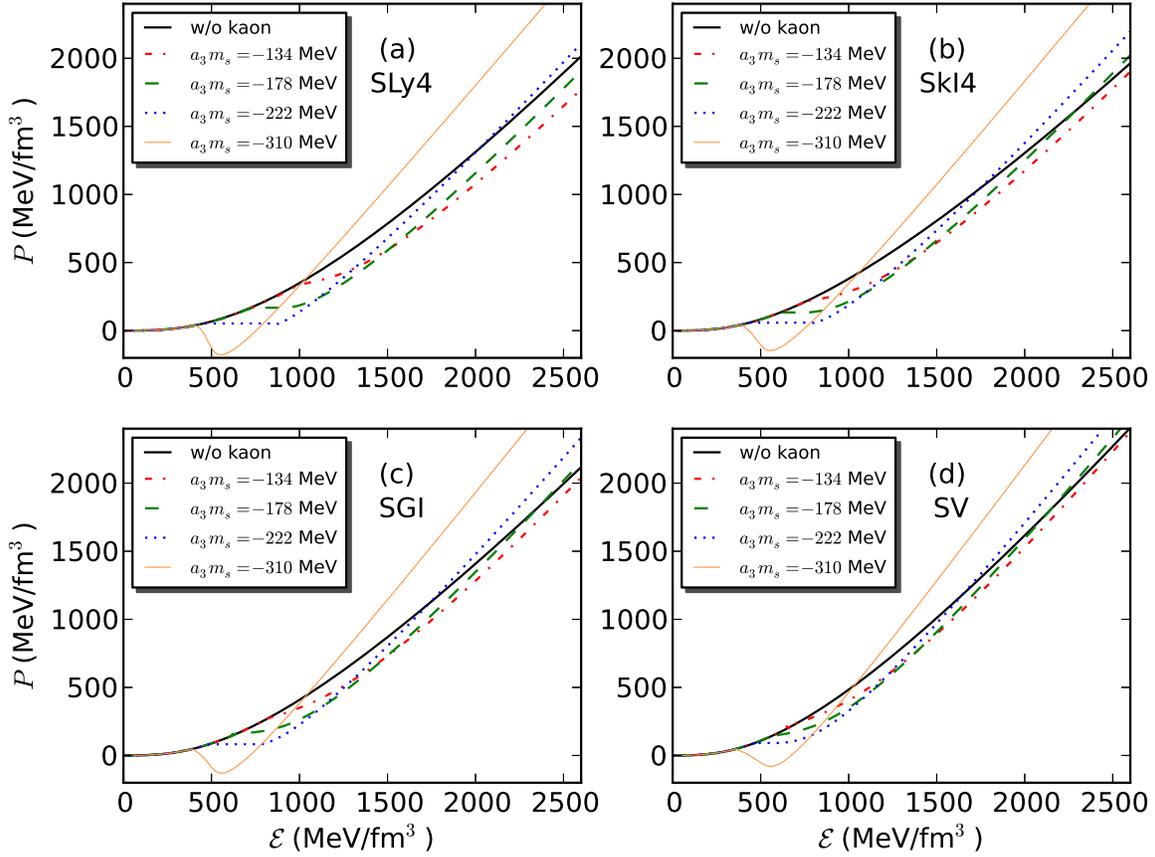}
\end{center}
\caption{Equation of state for each of four SHF models (color figures are available in online version). 
For $a_3 m_s=-178$ and $-222$ MeV, Maxwell construction is used to make the pressure constant 
for the unstable energy density region. In case of $a_3m_s=-310$ MeV, Maxwell
construction is not possible since the kaon condensation makes the system too soft and 
thermodynamically unstable for the low energy density region, hence the neutron star has finite surface density.
}
\label{fig:eos}
\end{figure}

\section{Neutron Stars with Kaon Condensation \label{sec3}}

\subsection{Equation of state and critical densities}

\begin{table}
\begin{center}
\begin{tabular}{lcccc}
\hline
Model \phantom{a} & $a_3m_s = -134$ MeV \phantom{a}
 & $a_3m_s = -178$ MeV  \phantom{a}
      & $a_3m_s = -222$ MeV \phantom{a}
      & $a_3m_s = -310$ MeV \\
\hline
SLy4 & 0.8580\, (5.36) & 0.6887\, (4.30) & 0.5689\, (3.56) & 0.4183\, (2.61) \\
SkI4  & 0.6813\, (4.26) & 0.5830\, (3.64) & 0.5070\, (3.17) & 0.3962\, (2.48) \\
SGI  & 0.7002\, (4.52) & 0.5890\, (3.80) & 0.5060\, (3.26) & 0.3921\, (2.53) \\
SV   & 0.5944\, (3.83) & 0.5139\, (3.32) & 0.4512\, (2.91) & 0.3608\, (2.33) \\
\hline
\end{tabular}
\end{center}
\caption{Critical densities in unit of fm$^{-3}$ (values in parentheses are in unit of $\rho_0$) for four SHF models 
with kaon condensation included. The critical density for kaon condensation decreases as $a_3m_s$ decreases, i.e., 
the strangeness content increases.  
}\label{tb:critical}
\end{table}

Equation of state of nuclear matter, which in general is the relation between pressure and
energy density, can be calculated from Eqs.~(\ref{eq:enden}) and (\ref{eq:pressure}). 
We employ four Skyrme force models (SLy4, SkI4, SGI, and SV) 
to calculate the uniform nuclear matter EoS ($\rho > 0.08$ fm$^{-3}$).
In the regime where the density is smaller than the density of uniform nuclear matter 
($\rho < 0.08$ fm$^{-3}$), we use liquid droplet model to treat heavy nuclei and free
gas of neutrons and electrons (see Appendix). Once we choose one specific SHF model 
for the EoS calculation of the entire neutron star, we apply the same model to both the uniform 
nuclear matter (high density regime) and the non-uniform lattice nuclear matter (low density regime). 


Fig.~\ref{fig:eos} shows the EoS for each of four models with four values of $a_3 m_s$ that we select. 
Note that the local charge neutrality is assumed for both cases with and without kaon condensation.
The results obtained without kaon condensation (black thick solid line in each panel) show that 
SLy4 is the softest while SV is the stiffest EoS, which can be expected from the values of $L$ and $K$ in Table~\ref{tab1}.
As the energy density increases from 0, kaon starts to condense at the density 
where the curves with kaon deviate from those without kaon. 
Table~\ref{tb:critical} summarizes the numerical results for the critical densities
at which the kaon condensation appears.
Numbers in the parentheses denote the critical densities in unit of $\rho_0$ for each model.
As indicated in the Fig.~\ref{fig:eos} and Table~\ref{tb:critical}, 
kaons condense earlier with smaller $a_3 m_s$ values (i.e., larger strangeness content) for all models.
Table~\ref{tb:critical} shows that the change of the critical densities 
resulting from change of $a_3 m_s$ is the largest for SLy4 which has the softest EoS among the four models. 
This implies that the softer nuclear models are more sensitive to the existence of kaons.
This sensitivity can also be deduced from the EoS in Fig.~\ref{fig:eos} by noting that
the width of band between the curves for $a_3 m_s=-134$~MeV and $-222$~MeV becomes
narrower with a stiffer nuclear EoS. 
%
In Fig.~\ref{fig:eos}, 
with $a_3 m_s=-134$~MeV, kaons soften the EoS in the entire energy density region.
However, for other values of $a_3 m_s$, kaons harden the EoS at very high energy densities.
%
Nonlinear chiral model for the meson-baryon interactions has been already employed in the
work by Thorsson et al. \cite{thorsson1994}. By assuming simple functional forms for the description
of nuclear forces and considering various compression modulus in the range of $K=120-240$~MeV, 
they obtained critical densities in the range of $\rho_{\rm crit} = (2.30-4.95) \rho_0$, which
are similar to what we obtain with more realistic nuclear models in this work.

After the formation of kaon condensation, we have an unstable part in the EoS 
where the derivative of pressure with respect to the energy density is negative. 
This unstable region can be treated by either
Maxwell construction or Gibbs condition.
In this work we adopt the Maxwell constructions, and the flat parts in the EoS 
are the consequences of the Maxwell constructions.
However, the Maxwell construction  is not
possible for $a_3 m_s=-310$~MeV 
because the pressure decreases so much that
the mean value of the pressure is negative for the low density region. 
Hence the neutron star has finite surface density and the resulting mass and radius of neutron star with $a_3 m_s=-310$ MeV are far from the current observations \cite{Demo10,Anto13}. 
Therefore, we do not include the results with $a_3 m_s = -310$ MeV in the later discussion. 

\subsection{Particle fractions}

\begin{figure}
\begin{center}
\includegraphics[scale=0.55]{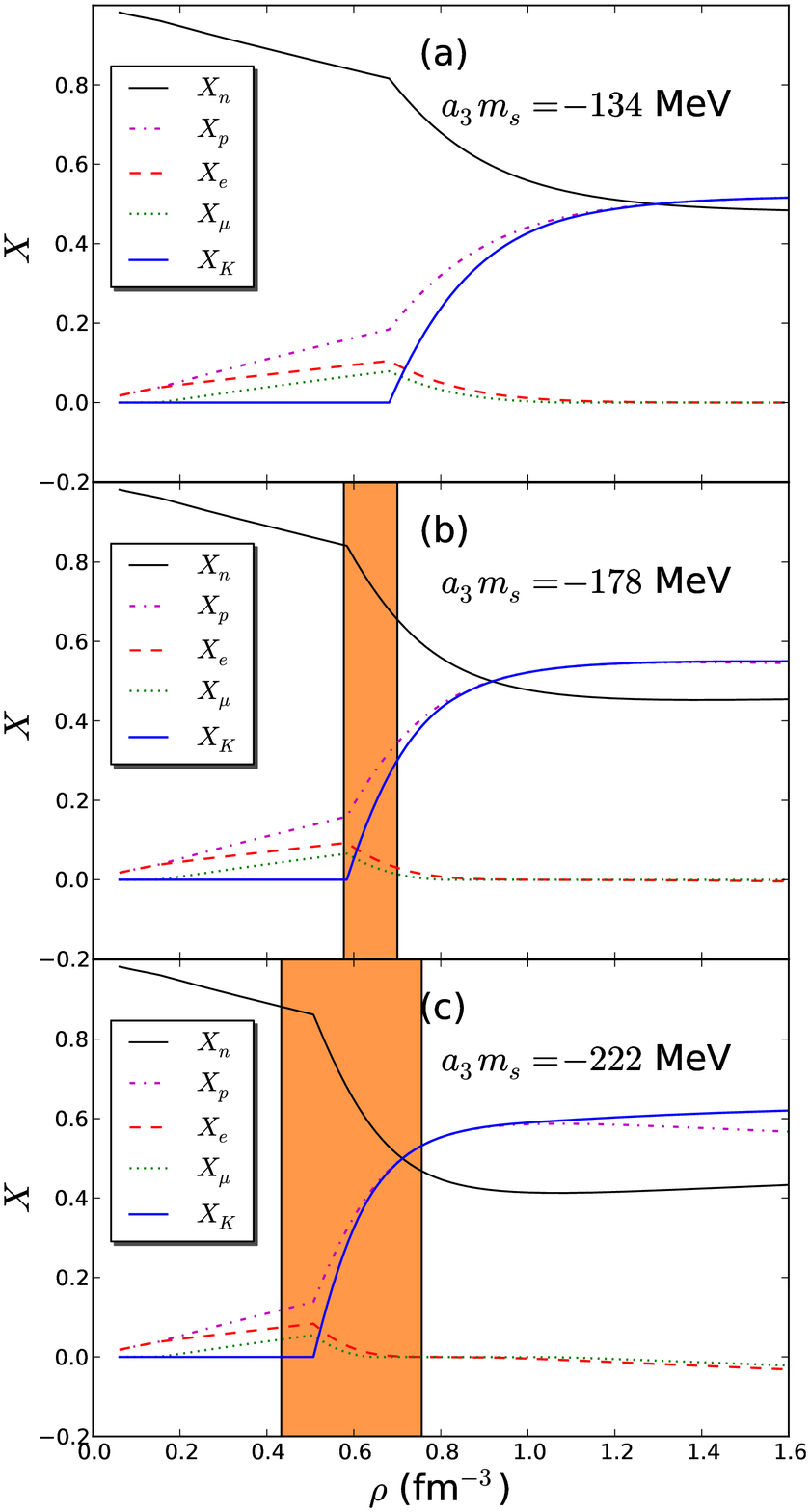}
\end{center}
\caption{Particle fractions of nuclear matter defined as the densities of each particle 
divided by baryon density for the SkI4 model with kaon condensation included
(color figures are available in online version). 
Negative fraction means anti-particle corresponding to each particle. 
Band intervals in (b) and (c) 
represent the Maxwell construction area where the pressure is constant.}
\label{fig:fraction}
\end{figure}

In Fig.~\ref{fig:fraction}, we show the particle fractions 
defined as the densities of each particle divided by the baryon density, for the SkI4 model.
Both kaon and proton densities increase very rapidly right after the critical densities due to the local charge neutrality.
Note that the local charge neutrality is implied in this figure and, even though we plot particle fractions for all densities, there exist density gaps in the interior of neutron stars due to the Maxwell construction.
%
In this figure, as $a_3 m_s$ decreases,
both proton and kaon fractions increase beyond neutron fraction,
which enhances the contribution of the symmetry energy to the EoS.
%
%
%
Unlike  leptons, kaon is not constrained by the Pauli blocking, 
and most of the leptons are suppressed by kaons at high densities making
%
the fraction of kaons almost equal to that of protons. 
%

\subsection{Mass and radius of neutron star}

The relation between mass and radius of cold neutron star can be obtained by solving 
the Tolman-Oppenheimer-Volkoff (TOV) equations numerically,
\begin{eqnarray}
\frac{dp}{dR} &=& - \frac{G(M(R)+4\pi R^{3} p/c^{2})(\epsilon + p)}
{R(R-2GM(R)/c^{2})c^{2}}, \\
\frac{dM}{dR} &=& 4\pi \frac{\epsilon}{c^{2}}R^{2},
\label{eq:tov}
\end{eqnarray}
where $\epsilon$ and $p$ denote the energy density and pressure, respectively.


\begin{figure}
\begin{center}
\includegraphics[scale=0.85]{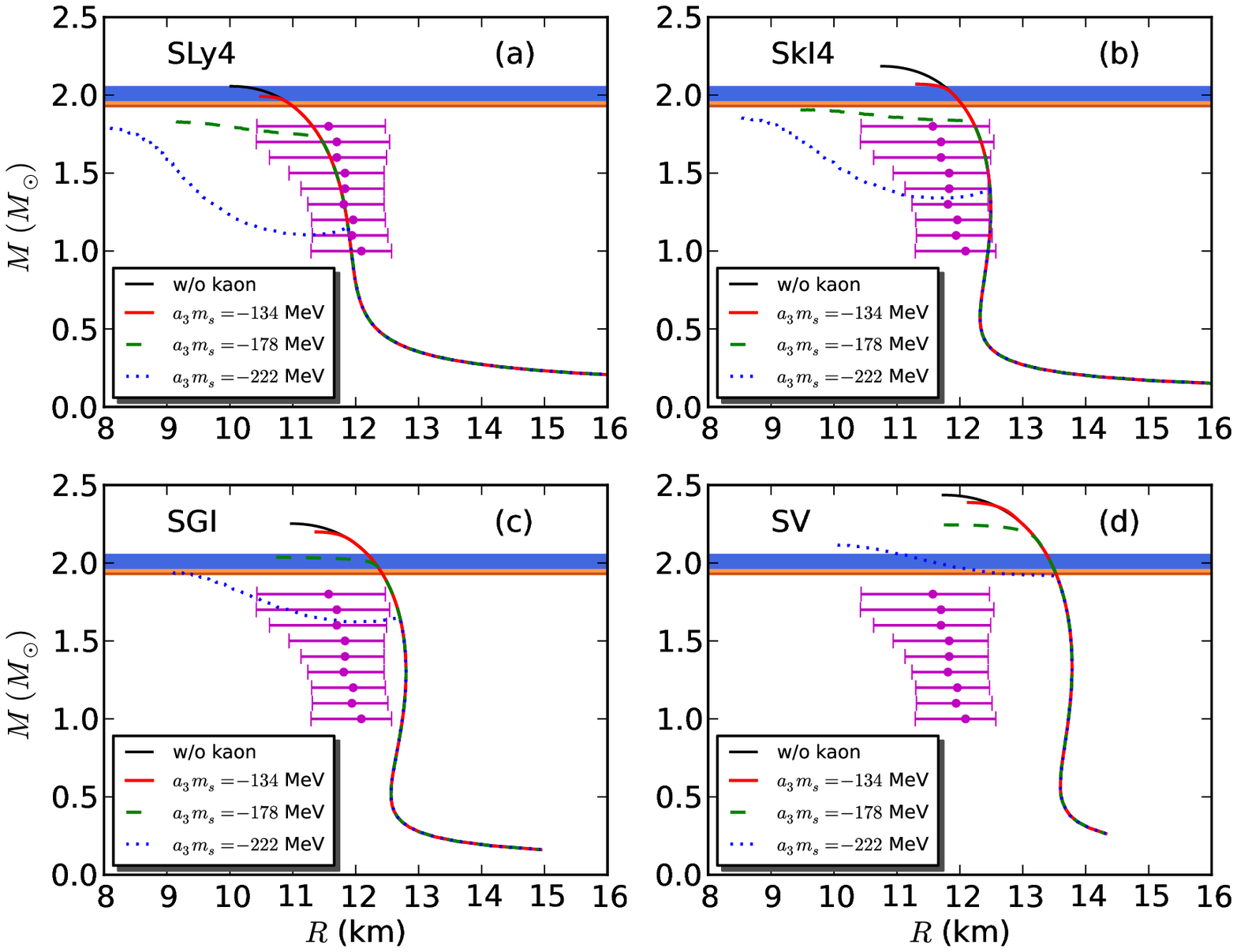}
\end{center}
\caption{Mass-radius curve for each of four SHF models (color figures are available in online version). 
Thick blue and thick orange solid straight lines are recent observations \cite{Demo10,Anto13}.
Filled circles with error bars denote
 the allowed mass and radius ranges obtained from the analysis by Steiner et al. \cite{slb2010}.
}
\label{fig:nsmr}
\end{figure}

In Fig.~\ref{fig:nsmr} we show the mass of neutron stars as a function of radius for four SHF models.
Thick blue and thick orange solid straight lines indicate the mass range 
of PSR J1614$-$2230 and J0348$+$0432, respectively \cite{Demo10,Anto13} 
and filled circles with error bars denote the allowed mass and radius ranges obtained from the analysis 
by Steiner et al. \cite{slb2010}.
Again, Fig.~\ref{fig:nsmr} confirms that the maximum mass increases as the stiffness of EoS increases.

Table~\ref{tb:kaonmass} summarizes the maximum mass of neutron stars predicted from four SHF models with 
kaon condensation included. For the comparison, the maximum mass obtained from the same models 
without kaon condensation (from Tab.~\ref{tab1}) is also shown in the last column of Tab.~\ref{tb:kaonmass}. 
For $a_3 m_s=-134$~MeV, the effect of kaon condensation on the maximum mass of neutron stars 
is rather weak, reducing the maximum mass by less than 4~\%.
With smaller $a_3 m_s$ values, the curves deviate more dramatically from those without kaons.
Note that in all models except SV, the existence of the unstable part with $a_3 m_s = -222$~MeV, 
which is near the kink at $R=11-12$ km having a positive
slope in the mass-radius curve (Fig.~\ref{fig:nsmr}), 
is an artifact of the Maxwell construction. 
%
%
%
%
With the Gibbs condition, the kink generally disappears and the curve becomes smooth 
(see e.g. \cite{ryujkps}). Since the Gibbs condition only affects the unstable region, 
our conclusion in this work remains unchanged.


In Fig.~\ref{fig:nsmr}, the results with $a_3 m_s=-222$ MeV are not consistent with 
either of the observed neutron star masses or the mass-radius ranges obtained 
by Steiner et al.\cite{slb2010}, 
regardless of any choice of the SHF models. In contrast, with $a_3 m_s=-134$ MeV, both SLy4 and SkI4 
are consistent with observations. 

Recenlty, Guillot {\it et al.} \cite{gswr2013} have measured small radii of neutron stars 
($R_{\infty} = 9.1_{-1.5}^{+1.3}$ km with the $90\%$ confidence level: see Ref.~\cite{wff1988},
however, for an alternative interpretation of this data)
using the thermal spectra from quiescent low-mass X-ray binaries inside five globular clusters. 
The smaller radii of neutron stars that they measured prefer the softer EoS such as 
Wiringa {\it et al.'s} \cite{wff1988} and rule out most of the presently popular EoS. 
It is interesting that our results with kaon condensation could explain the existence 
of such a neutron star that has a radius of $\sim 9$ km and a mass of $\sim 2.0~M_{\odot}$ 
(e.g., see the $a_3 m_s=-222$ MeV case of SGI in the panel (c) of Fig. 3). 

\begin{table}
\caption{Maximum mass of neutron star (in unit of $M_{\odot}$)
 in the presence of kaon for each model.}
\begin{center}
\begin{tabular}{lcccc}
\hline
Model \phantom{a} & $a_3m_s=-134$ MeV \phantom{a}
& $a_3m_s=-178$ MeV\phantom{a} & $a_3m_s=-222$ MeV & ~~~~~~no kaon ~~~~~~\\
\hline
SLy4 & 1.99 & 1.83 & 1.79 & 2.07 \\
SkI4 & 2.07 & 1.91 & 1.85 & 2.19 \\
SGI  & 2.20 & 2.04 & 1.94 & 2.25 \\
SV   & 2.39 & 2.24 & 2.12 & 2.44 \\
\hline
\end{tabular}
\end{center}
\label{tb:kaonmass}
\end{table}

\section{Conclusion \label{sec4}}

This work is motivated by the fact that many SHF models,
which are excellent in reproducing the properties of known nuclei, are inconsistent in predicting 
nuclear matter properties at supra-nuclear densities. 
We select four SHF models that are consistent with recent observations 
of neutron star masses \cite{Demo10,Anto13} 
in order to investigate and understand the effect of kaon condensation.
Since the interactions of kaon in nuclear medium are quite uncertain, 
we employ four different parameter sets to cover a wide range of kaon interactions.


As one can see in Fig.~\ref{fig:nsmr}, SLy4 and SkI4  are consistent 
with the recent constraints \cite{Demo10,Anto13,slb2010} even without kaon condensation. 
Adding kaons to these models, the mass-radius relation with kaon deviates from the those
without kaons more drastically for smaller $a_3 m_s$ values.
As a result, the behavior with $a_3 m_s = -222$~MeV satisfies the mass-radius relation
constraints in a limited manner.
This result shows that the observation of the neutron star can provide
constraints on the strangeness content of the proton, and our result implies that 
kaon condensation, if it occurs, should occur only for heavy neutron stars ($>1.9 M_\odot$), 
with $a_3 m_s$ being greater than $-178$ MeV (or the strangeness content being smaller than $0.05$).
This small allowed value of strangeness content of the proton is in accordance with 
experiments \cite{arm2012}, lattice calculation \cite{doi2009}, and a recent updated calculation
with a chiral effective theory \cite{wang2013}.
This would be also consistent with the recent observation of Cas A and its cooling simulation \cite{dpls2011} because 
kaon condensation does not affect thermal evolution of 
neutron stars with smaller mass ($M < 1.9 M_{\odot}$). 


We have discussed the contribution of strangeness by focusing only on the kaon in this work, but
we have to consider hyperons as well for the consistency. 
In the current SHF approach, to our knowledge, 
there are more than 10 models for hyperon-nucleon interactions and 
3 models for the hyperon-hyperon interactions in nuclear matter. 
Our recent analysis shows that the EoS at high densities and the resulting mass and radius of 
neutron star are also highly sensitive to the hyperon-nucleon and hyperon-hyperon interactions in
nuclear medium. We will present the work with hyperons in the future. 

\section*{Acknowledgments}
We would like to thank the referee for the valuable criticisms and comments.
CHH is grateful to E. Hiyama for useful comments about the recent results
in the hypernuclear interactions. 
Work of CHH and YL was supported by the Basic Science Research 
Program through the National Research Foundation of Korea (NRF)
funded by the Ministry of Education, Science and Technology 
Grant No. 2010-0023661. 
YL was also supported by
the Rare Isotope Science Project funded by the Ministry of Science, ICT and
Future Planning (MSIP) and National Research Foundation (NRF) of Korea.
CHL was supported by the BAERI Nuclear R \& D program (M20808740002) of 
MEST/KOSEF and the Financial Supporting Project of Long-term Overseas 
Dispatch of PNU's Tenure-track Faculty, 2013. KK was supported by the year 
2013 Research Fund of the Ulsan National Institute of Science and Technology 
(UNIST). 

\appendix
\label{sec:appendix}
\section{Heavy nuclei with dripped neutrons} 
We follow the discussion of Lattimer \& Swesty \cite{ls1991}.
The total energy density (without electron contribution) is given by
\begin{equation}\label{eq:heavy}
F = \unifi + \frac{3\su}{\rn} \left [ \sigx +\mus\nun \right ] 
+ \frac{4\pi}{5}(\rn\nii\xii e)^2 \cu + (1-u)\nno\foo\,\,,
\end{equation}
where $u$ is a volume fraction of heavy nuclei to Wigner-Seitz cell,
$n_i$ is the density of heavy nuclei inside, $f_i$ is 
the energy per baryon
of the heavy nuclei, $s(u)$ is surface shape factor, 
$r_N$ is the radius of heavy nuclei,
$\sigma(x)$ 
is a surface tension as a function of proton fraction $x$,
$\mu_s$ is the neutron chemical potential on the surface,
$\nu_n$ is the areal neutron density on the surface, 
$x_i$ is the proton fraction of heavy nuclei,
$c(u)$ is the Coulomb shape function, 
$n_{no}$ is the neutron density outside of heavy nuclei,
and $f_o$ is the energy per baryon outside of the heavy nuclei. 

We employ Lagrange-Multiplier
method with constraints (baryon number density and charge neutrality),
thus for given baryon number density ($n$) and proton fraction ($Y_p$),
we have
\begin{equation}
G = F + \la \left [ n - \uni -3\su \frac{\nun}{\rn} -(1-u)\nno \right ] + \lb(nY_{p} -\unixi)\,.
\end{equation}
In the above equations, unknowns are  $\nii$, $x_i$, $r_N$, $x$, $\frac{\nun}{\rn}$,
$u$, and $n_{no}$.
\begin{align}
\frac{\pd G}{\pd \nii} = 0 : & \quad 
u(\muni -\xii\muhi) + \frac{8\pi}{5}(\rn \xii e)^2 \nii \cu
-\la u -\lb u \xii =0,
\label{eq:dgdni2}
\\
\frac{\pd G}{\pd \xii} = 0 : & \quad 
-u \nii \muhi + \frac{8\pi}{5}(\rn \nii e)^2 \xii \cu 
-\lb \uni =0,
\label{eq:dgdxi2}
\\
\frac{\pd G}{\pd \rn} = 0 : & \quad
-\frac{3\su}{\rn^2}\sigma + \frac{8\pi}{5}(\nii \xii e)^2 \cu =0,
 \label{eq:dgdrn2}
\\
\frac{\pd G}{\pd x} = 0 : & \quad \frac{3\su}{\rn}
\left(\frac{\pd \sigma}{\pd x} + \nun \frac{\pd \mus}{\pd x}\right) =0,
\label{eq:dgdx2}
\\
\frac{\pd G}{\pd (\nun/\rn)} = 0 : & \quad 3\su(\mus -\la) =0,
\label{eq:dgdnun2}
\\
\frac{\pd G}{\pd u} = 0 : & \quad 
 \nii \fii +\frac{3\spr}{\rn}(\sigma + \mus\nun)
 +\frac{4\pi}{5}(\rn\nii\xii e)^2\cpr 
 -\la(\nii + 3\spr \frac{\nun}{\rn} -\nno) -\lb \nii \xii =0,
\label{eq:dgdu2}
\\
\frac{\pd G}{\pd \nno} = 0 : & \quad
(1-u)\muno - (1-u)\la =0,
\label{eq:dgdnno}
\\
\frac{\pd G}{\pd \la} = 0 : & \quad
\nii - \uni -3\su \frac{\nun}{\rn} =0,
\label{eq:dgdla2}
\\
\frac{\pd G}{\pd \lb}= 0 : & \quad 
nY_{p} - \unixi =0\,,
\label{eq:dgdlb2}
\end{align}
where $\hat{\mu}_i = \mu_{ni} - \mu_{pi}$.
$\mu_{ni}$ and $\mu_{pi}$ are neutron and proton chemical potential 
inside of the heavy nuclei, respectively.\\
Eq. \eqref{eq:dgdni2}$\times$ $\nii -$ \eqref{eq:dgdxi2}$\times$$\xii$ gives
$\la = \muni$.
From Eq. \eqref{eq:dgdnun2}, we have $ \la = \mus$. 
From Eq. \eqref{eq:dgdnno}, $\la = \muno$, so $\muni=\muno=\mus$. \\
Similarly we have $\lb$ from Eq. \eqref{eq:dgdxi2},
\begin{equation}
\begin{aligned}
\lb  & = -\muhi + \frac{1}{\uni}\frac{8\pi}{5}(\rn \nii e)^2 \xii \cu \\
     & = -\muhi + \frac{1}{\unixi}\frac{2}{3}\beta \mathcal{D}(u)
     \label{eq:lambda2b}\,,
\end{aligned}
\end{equation}
where $\beta=9(\frac{\pi e^2x_i^2 n_i^2 \sigma^2}{15})^{1/3}$, and
 $\mathcal{D}=[c(u)s^{2}(u)]^{1/3}$ is a geometric shape
function which corresponds to nuclear pasta phase in liquid droplet model
 \cite{ls1991}.
\\
Finally, if we plug Eq. \eqref{eq:lambda2b} into Eq. \eqref{eq:dgdu2} then we get,
\begin{equation}
\begin{aligned}
\nii \fii & + \beta \mathcal{D}^{\prime} -\muni \nii + \nii \xii \muhi
-\frac{2}{3u}\beta{\mathcal{D}} +\muno\nno -\nno\foo  =0 \\
\Rightarrow P_i & -P_o - \beta\left(\mathcal{D}^{\prime} -\frac{2\mathcal D}{3u} \right) =0,
\end{aligned}\
\end{equation}
where $P_i$ ($P_o$) is pressure inside (outside) of the heavy nuclei.
Thus, we have four equations to solve 
\begin{equation}
\begin{aligned}
&P_i  -P_o - \beta\left(\mathcal{D}^{\prime} -\frac{2\mathcal D}{3u} \right) =0, \\
&\unixi - n Y_{p} = 0, \\
&\uni + \frac{2\beta}{3\sigma}\cald\nun + (1-u)\nno - n =0 ,\\
&\muni -\muno = 0,
\end{aligned}
\end{equation}
with four unknowns, $u$ ( or $\ln u$ ), $\nii$, $\nno$ ( or $\ln \nno$ ), and $\xii$.

Thermodynamic quantities for this case is given by the same formalism with
hot dense matter but without alpha particle and translational term, so
\begin{equation}
\begin{aligned}
\muh  & = \muhi -\frac{2}{3\unixi}\beta\cald, \\
\mun & =\muno, \\
P & = P_o -\beta(\cald -u\caldp)\,.
\end{aligned}
\end{equation}
In case of neutron star outer crust, we can construct eq. \eqref{eq:heavy}
without $\nno$ and follow the same methodology with the case of
neutron star inner crust.


\begin{thebibliography}{5}

\bibitem{brown2000}
B.~A. Brown, Phys. Rev. Lett. {\bf 85}, 5296 (2000).

\bibitem{thesis2012} `PhD thesis', Yeunhwan Lim (2012).

\bibitem{mornas05}
L. Mornas, Eur. Phys. J. A {\bf 24}, 293 (2005).

\bibitem{guleria12}
N. Guleria, S.~K. Dhiman, and R. Shyam,
Nucl. Phys. A {\bf 886}, 71 (2012).

\bibitem{ryu11}
C.-Y. Ryu, C.~H. Hyun, C.-H. Lee,
Phys. Rev. C {\bf 84}, 035809 (2011).

\bibitem{Demo10}
P. Demorest, T. Pennucci, S. Ransom, M. Roberts, and J.~W.~T. Hessels,
Nature {\bf 467}, 1081 (2010).

\bibitem{Anto13}
J. Antoniadis {\it et al.}, Science {\bf 340}, 448 (2013).

\bibitem{ryu07}
C.~Y. Ryu, C.~H. Hyun, S.~W. Hong, and B.~T. Kim,
Phys. Rev. C {\bf 75}, 055804 (2007).

\bibitem{sple2005} 
A.~W. Steiner, M. Prakash, J.~M. Lattimer, and P.~J. Ellis, 
Phys. Rep. {\bf 411}, 325 (2005).

\bibitem{klmn2010} 
M. Kortelainen,  T. Lesinski, J. More , W. Nazarewicz, J. Sarich,  
N. Schunck, M.~V. Stoitsov, and S. Wild,
Phys. Rev. C {\bf 82}, 024313 (2010).


\bibitem{knorren95}
R. Knorren, and M. Prakash, and P.~J. Ellis
Phys. Rev. C {\bf 52}, 3470 (1995).
\bibitem{kaplan86}
D.~B. Kaplan, and A.~E. Nelson, Phys. Lett. B {\bf 175}, 57 (1986).
\bibitem{thorsson94}
B. Thorsson, M. Prakash, and J.~M. Lattimer,
Nucl. Phys. A {\bf 572}, 693 (1994).
\bibitem{baym73}
G. Baym, Phys. Rev. Lett. {\bf 30}, 1340 (1973).
\bibitem{thorsson1994}
V. Thorsson, M. Prakash, and J.~M. Lattimer,
Nucl. Phys. A {\bf 572}, 693 (1994).
\bibitem{slb2010} 
A.~W. Steiner, J.~M. Lattimer, and E.~F. Brown, 
ApJ {\bf 722}, 33 (2010).
\bibitem{ryujkps}
C.~Y. Ryu, C.~H. Hyun, and S.~W. Hong,
J. Korean Phys. Soc. {\bf 54}, 1448 (2009).

\bibitem{gswr2013}
S. Guillot, M. Servillat, N.~A. Webb, and R.~E. Rutledge,
ApJ {\bf 772}, 7 (2013).
\bibitem{wff1988}
R.~B. Wiringa, V. Fiks, and A. Fabrocini,
Phys. Rev. C {\bf 38}, 1010 (1988).

\bibitem{arm2012}
D.~S. Armstrong, and R.~D. McKweon,
Ann. Rev. Nucl. Part. Sci, {\bf 62}, 337 (2012).
\bibitem{doi2009}
T. Doi, M. Deka, S.-J. Dong, T. Draper, K.-F. Liu, D. Mankame, N. Mathur,
and T. Streuer,
Phys. Rev. D {\bf 80}, 094503 (2009).
\bibitem{wang2013}
P. Wang, D.~B. Leinweber, and A.~W. Thomas,
arXiv:1312.3375 [hep-ph].
\bibitem{dpls2011}
D. Page, M. Prakash, J.~M. Lattimer, and A.~W. Steiner,
Phys. Rev. Lett.{\bf 106}, 081101 (2011).




\bibitem{ls1991}
J.~M. Lattimer and F.~D. Swesty, Nucl. Phys. A {\bf 535}, 331 (1991).



\end{thebibliography}
\end{document}